\documentstyle[12pt,epsfig]{article}

\begin{document}

\baselineskip=6.8mm

\def\ap#1#2#3{           {\it Ann. Phys. (NY) }{\bf #1} (19#2) #3}
\def\arnps#1#2#3{        {\it Ann. Rev. Nucl. Part. Sci. }{\bf #1} (19#2) #3}
\def\cnpp#1#2#3{        {\it Comm. Nucl. Part. Phys. }{\bf #1} (19#2) #3}
\def\apj#1#2#3{          {\it Astrophys. J. }{\bf #1} (19#2) #3}
\def\asr#1#2#3{          {\it Astrophys. Space Rev. }{\bf #1} (19#2) #3}
\def\ass#1#2#3{          {\it Astrophys. Space Sci. }{\bf #1} (19#2) #3}

\def\apjl#1#2#3{         {\it Astrophys. J. Lett. }{\bf #1} (19#2) #3}
\def\ass#1#2#3{          {\it Astrophys. Space Sci. }{\bf #1} (19#2) #3}
\def\jel#1#2#3{         {\it Journal Europhys. Lett. }{\bf #1} (19#2) #3}

\def\ib#1#2#3{           {\it ibid. }{\bf #1} (19#2) #3}
\def\nat#1#2#3{          {\it Nature }{\bf #1} (19#2) #3}
\def\nps#1#2#3{          {\it Nucl. Phys. B (Proc. Suppl.) }
                         {\bf #1} (19#2) #3}
\def\np#1#2#3{           {\it Nucl. Phys. }{\bf #1} (19#2) #3}
\def\pl#1#2#3{           {\it Phys. Lett. }{\bf #1} (19#2) #3}
\def\pr#1#2#3{           {\it Phys. Rev. }{\bf #1} (19#2) #3}
\def\prep#1#2#3{         {\it Phys. Rep. }{\bf #1} (19#2) #3}
\def\prl#1#2#3{          {\it Phys. Rev. Lett. }{\bf #1} (19#2) #3}
\def\pw#1#2#3{          {\it Particle World }{\bf #1} (19#2) #3}
\def\ptp#1#2#3{          {\it Prog. Theor. Phys. }{\bf #1} (19#2) #3}
\def\jppnp#1#2#3{         {\it J. Prog. Part. Nucl. Phys. }{\bf #1} (19#2) =
#3}

\def\rpp#1#2#3{         {\it Rep. on Prog. in Phys. }{\bf #1} (19#2) #3}
\def\ptps#1#2#3{         {\it Prog. Theor. Phys. Suppl. }{\bf #1} (19#2) #3=
}
\def\rmp#1#2#3{          {\it Rev. Mod. Phys. }{\bf #1} (19#2) #3}
\def\zp#1#2#3{           {\it Zeit. fur Physik }{\bf #1} (19#2) #3}
\def\fp#1#2#3{           {\it Fortschr. Phys. }{\bf #1} (19#2) #3}
\def\Zp#1#2#3{           {\it Z. Physik }{\bf #1} (19#2) #3}
\def\Sci#1#2#3{          {\it Science }{\bf #1} (19#2) #3}
\def\n.c.#1#2#3{         {\it Nuovo Cim. }{\bf #1} (19#2) #3}
\def\r.n.c.#1#2#3{       {\it Riv. del Nuovo Cim. }{\bf #1} (19#2) #3}
\def\sjnp#1#2#3{         {\it Sov. J. Nucl. Phys. }{\bf #1} (19#2) #3}
\def\yf#1#2#3{           {\it Yad. Fiz. }{\bf #1} (19#2) #3}
\def\zetf#1#2#3{         {\it Z. Eksp. Teor. Fiz. }{\bf #1} (19#2) #3}
\def\zetfpr#1#2#3{     {\it Z. Eksp. Teor. Fiz. Pisma. Red. }{\bf #1} (19#2=
) #3}
\def\jetp#1#2#3{         {\it JETP }{\bf #1} (19#2) #3}
\def\mpl#1#2#3{          {\it Mod. Phys. Lett. }{\bf #1} (19#2) #3}
\def\ufn#1#2#3{          {\it Usp. Fiz. Naut. }{\bf #1} (19#2) #3}
\def\sp#1#2#3{           {\it Sov. Phys.-Usp.}{\bf #1} (19#2) #3}
\def\ppnp#1#2#3{           {\it Prog. Part. Nucl. Phys. }{\bf #1} (19#2) #3=
}
\def\cnpp#1#2#3{           {\it Comm. Nucl. Part. Phys. }{\bf #1} (19#2) #3=
}
\def\ijmp#1#2#3{           {\it Int. J. Mod. Phys. }{\bf #1} (19#2) #3}
\def\tp{these proceedings}
\def\pc{private communication}
\def\ip{in preparation}
\relax

\newcommand{\GeV}{\,{\rm GeV}}
\newcommand{\MeV}{\,{\rm MeV}}
\newcommand{\keV}{\,{\rm keV}}
\newcommand{\eV}{\,{\rm eV}}
\newcommand{\Tr}{{\rm Tr}\!}
\renewcommand{\arraystretch}{1.2}
\newcommand{\be}{\begin{equation}}
\newcommand{\ee}{\end{equation}}
\newcommand{\beqa}{\begin{eqnarray}}
\newcommand{\eeqa}{\end{eqnarray}}
\newcommand{\ba}{\begin{array}}
\newcommand{\ea}{\end{array}}
\newcommand{\bmat}{\left(\ba}
\newcommand{\emat}{\ea\right)}
\newcommand{\refs}[1]{(\ref{#1})}
\newcommand{\ler}{\stackrel{\scriptstyle <}{\scriptstyle\sim}}
\newcommand{\ger}{\stackrel{\scriptstyle >}{\scriptstyle\sim}}
\newcommand{\lag}{\langle}
\newcommand{\rag}{\rangle}
\newcommand{\ns}{\normalsize}
\newcommand{\cm}{{\cal M}}
\newcommand{\gr}{m_{3/2}}
\newcommand{\p}{\partial}

\def\rp{ $R_P$}
\def\321{$SU(3)\times SU(2)\times U(1)$}
\def\tl{{\tilde{l}}}
\def\tL{{\tilde{L}}}
\def\bd{{\overline{d}}}
\def\tL{{\tilde{L}}}
\def\a{\alpha}
\def\b{\beta}
\def\g{\gamma}
\def\c{\chi}
\def\d{\delta}
\def\D{\Delta}
\def\db{{\overline{\delta}}}
\def\Db{{\overline{\Delta}}}
\def\e{\epsilon}
\def\l{\lambda}
\def\n{\nu}
\def\m{\mu}
\def\nt{{\tilde{\nu}}}
\def\p{\phi}
\def\P{\Phi}
\def\x{\xi}
\def\r{\rho}
\def\s{\sigma}
\def\t{\tau}
\def\th{\theta}
\def\ne{\nu_e}
\def\nm{\nu_{\mu}}
\def\rp{$R_P$}
\def\mp{$M_P$}    
\renewcommand{\Huge}{\Large}
\renewcommand{\LARGE}{\Large}
\renewcommand{\Large}{\large}

\begin{flushright}
hep-ph/9904211\\
IC/99/32\\
NYU-TH/99/3/03/
\end{flushright}
\vskip 2.0truecm
\begin{center}

{\Large \bf Probing Large Extra  Dimensions with Neutrinos}

\vskip 2.0truecm

{ Gia ~Dvali$^{1,2}$  and Alexei Yu. Smirnov$^{2, 3}$}\\
\vskip 0.5truecm

{\ns \it $^{1)}$ Physics Department, 
New York University, New York, NY 10003, USA}\\

\vskip 0.2truecm

{\ns \it $^{2)}$ Abdus Salam International Centre for Theoretical Physics,
I-34100 Trieste, Italy}\\

\vskip 0.2truecm

{\ns \it $^{3)}$ Inst. for Nucl. Research, Russian Academy of Sciences,
107370 Moscow, Russia}

\vspace{2cm}

\begin{abstract}
{\ns
We study  implications of theories with 
sub-millimeter extra dimensions
and  $M_f \sim (1 - 10) $ TeV scale quantum gravity for  neutrino
physics.  
In these theories, the left-handed neutrinos as well as other standard
model (SM) particles, are localized on a brane embedded in the bulk of
large extra space.  Mixing of neutrinos with (SM) singlet fermions
propagating in the bulk  is naturally suppressed
by the volume factor $M_f/M_P  \sim 3\cdot 10^{-16} -  3\cdot 10^{-15}$,
where $M_P$ is the Planck mass. 
Properties  of the neutrino oscillations and the resonance
conversion to the bulk fermions are considered. 
We show that the resonance conversion of the
electron neutrinos to the light bulk fermions can solve 
the solar neutrino problem. The signature 
of the solution is the peculiar distortion of the solar  neutrino spectrum. 
The  solution implies
that the radius of at least one  extra dimension should be in the range
0.06 - 0.1 mm  {\it irrespective} of  total number 
of extra dimensions. 
The corresponding modification of the Newtonian law is within the range 
of sensitivity 
of proposed sub-millimeter experiments, thus  providing a verifiable link
between neutrino physics and the gravity measurements.}
\end{abstract}

\end{center}

\thispagestyle{empty}

\newpage

\section{Introduction}

It has been suggested recently \cite{add,aadd, add1} that the fundamental
scale of quantum gravity, $M_f$,  can be as low as few TeV. The observed
weakness
of gravity is the result of $N$ ( $\geq 2$) new space dimensions in
which gravity can propagate
\footnote{In a different
context an attempt of lowering the string scale to TeV, without lowering
the fundamental Planck scale was considered in \cite{lyk}, based on an
earlier observation in \cite{witten}, see also 
\cite{dienes} for lowering the GUT scale. Dynamical localization of
the fields
on a (solitonic) brane embedded in a higher dimensional space-time has
been suggested earlier in the field theoretic context
\cite{localization1/2}, \cite{localization1},\cite{ds}. For
some realizations of this scenario in the $D$-brane context see,
\cite{aadd}, \cite{bw}.}. 
The observed (reduced) Planck scale, $M_{P} = (4\pi G_N)^{-1/2} 
= 3.4 \cdot 10^{18}$ GeV, where $G_N$ is the Newton constant,   is
then related to the reduced Planck scale in $4 + N$ dimensions, 
$M_f$ (fundamental scale),  by 
\be
M_{P} = M_f \sqrt{ M_f^N V_N}~,
\label{relation}
\ee
where $V_N \equiv L_1 L_2....  L_N$ is the volume of the extra space,
and $L_i$ is the size of the $i$ - compact dimension.   
For definiteness we will assume that the volume has a configuration of
torus in which case $L_i = 2 \pi R_i$, where $R_i$  $(i = 1, 2,   ... N)$
are the radii of extra dimensions, so that  
\be
V_N = (2\pi)^N R_1  R_2  ...  R_N~. 
\label{volume}
\ee
Using (\ref{relation}) and  (\ref{volume}) we get 
the constraint on the extra dimension radii:  
\be
(2\pi)^N R_1 R_2 ... R_N = \frac{M_P^2}{M_f^{N +2}}~. 
\label{radii}
\ee
(Notice that in some publications the factor $(2\pi)^N$ is removed 
from this relation by redefinition of
the fundamental scale: $M_* = (2\pi)^{N/(N + 2)} M_f$.)
 
The phenomenological acceptance requires that $N \geq 2$, 
since for $N = 1$ the radius would be of the solar system size.
According to present measurement the distance above which the Newtonian
law should not be changed is about 1 mm, and therefore  
\be
L_i = 2\pi R_i \leq 1~ {\rm mm}~. 
\label{mmconstraint}
\ee
For $N = 2$ and $M_f \sim$ TeV one gets from (\ref{relation},\ref{volume}) 
$R_1 \sim R_2 \sim 0.1$ mm which satisfies (\ref{mmconstraint}) 

Thus, in theories under consideration it is expected that the Newtonian
$1/r$ law breaks down at the scales smaller than the largest
extra dimension: $L_{max}$. The experimentally most exiting possibility
would be if $L_{max} \sim 1 - 10^{-2}$ mm, that is,  in the range of
sensitivity of proposed experiments \cite{measurements}. 
As we will argue in this paper the same range 
is suggested by neutrino physics,  namely, by  a solution of the
solar neutrino problem based on  existence of
new dimensions. 

Usually it is assumed that 
all large radii $R_i$ are of the same order of
magnitude. In such a case $ N > 2$ would be well out of sensitivity of any
planned sub-millimeter gravitational measurements.   
On the other hand, for $N = 2$ 
the supernova analysis  pushes the lower 
bound on $M_f$ to  $30$ TeV 
\cite{add1} or even to
$50$ TeV \cite{supernova22}  implying that $R < 0.01$ mm, which is
again beyond the planned experimental sensitivity. 
However, in the absence of any commonly accepted mechanism for
stabilization of  large radii\footnote{For some ideas in this direction
see \cite{nimaetal}.},
the requirement of their equality is
unjustified. In the present paper we will 
assume that radii may take arbitrary values which satisfy 
the fixed over-all volume (\ref{radii}) and  phenomenological 
(\ref{mmconstraint}) constraints. In such a case the 
theory with several extra dimensions  
still can be subject of sub-millimeter test,  while  avoiding
astrophysical and other laboratory bounds. 
As we will see, these bounds are sensitive to the shape of extra dimensions.

In this paper we will discuss possible consequences 
of the high-dimensional theories for neutrino physics. 
In particular, we will suggest  new 
high-dimensional solution of the solar neutrino puzzle.  
This solution 
implies that the radius of at least one extra dimension must be
within $0.06  - 0.1 $ mm range. 
This observation relies on new  high dimension mechanism of neutrino mass
generation suggested in \cite{addm} which we will  briefly recall.

According to the  framework elaborated in  \cite{add,aadd, add1}, 
all the standard model particles must be localized on a
3-dimensional  hyper-surface 
('brane') \cite{localization1/2,localization1,ds}
embedded in the bulk of $N$ large extra dimensions.
The same is true for {\it any other} state charged
under the standard model group. The argument is due to the
conservation of the gauge flux, which indicates that no state carrying
a charge under a gauge field localized on the brane, can exist
away from it\cite{ds}\cite{add}.  
Thus, all the particles split in two categories:
those that live on the brane,  and those
which exist everywhere, 'bulk modes'. Graviton belongs to the
second category. Besides the graviton there can be  additional
neutral states propagating in the bulk. In general, the couplings between
the brane, $\psi_{brane}$, 
and the bulk $\psi_{bulk}$  
modes are suppressed by a volume factor:
\be
\frac{1 }{ \sqrt{M_f^N V_N}} \psi_{brane} \psi_{brane}
\psi_{bulk}~. 
\label{coupl}
\ee
According to (\ref{relation}) the coupling constant in (\ref{coupl}) 
equals  
\be  
 \frac{M_f}{M_P} =
3 \cdot 10^{-16}\frac{M_f}{ 1 {\rm TeV}}
\ee
and it does not depend on  number of extra dimensions. 
It was suggested \cite{addm} to use this small
model-independent coupling to explain the smallness of the neutrino mass. 
The left handed neutrino, $\nu_L$, having weak isospin and
hypercharge must reside on the brane. Thus, it can get a
naturally small Dirac mass through the mixing with some bulk fermion
and the latter can be interpreted as the right-handed neutrino
$\nu_R$:
\be
\frac{h M_f}{M_P} H \bar{\nu}_L \nu_R~. 
\label{inter}
\ee
Here $H$ is the Higgs doublet and $h$ is the model-dependent Yukawa
coupling.
After electro weak symmetry breaking the interaction (\ref{inter}) will
generate
the mixing mass
\be
m_D  =  \frac{h v M_f}{M_P}~, 
\label{Dmass}
\ee
where $v$ is the VEV of $H$. For $M_f \sim 1$ TeV
and $h = 1$ this  mass is about  $5.6 \cdot 10^{-5}$ eV.

Being the bulk state,  $\nu_R$ has a whole tower of
the Kaluza-Klein (KK) relatives. 
For $N$ extra dimensions they can be labeled by a set of
$N$ integers ${n_1, n_2 ... n_N}$ (which determine momenta in extra 
spaces): 
 $\nu_{n_1, n_2 ... n_N ~R}$.
Masses of these states are given by:
\be
m_{n_1, n_2 ... n_N} = \sqrt{\sum_i \frac{n_i^2}{R_{i}^2}}~. 
\ee
Notice that the masses of the KK states are determined by the {\it radii}  
whereas the scale of the Newton law modification 
is given by the {\it size} of  compact dimensions.

The left handed neutrino couples  with all
$\nu_{n R}$ with the same mixing mass (\ref{Dmass}).
This mixing is possible due to the spontaneous breaking of the
translational invariance in the bulk by the brane. 

In  ref. \cite{ddg} a general case  has been
considered with possible universal Majorana mass 
terms for the bulk fermions. 
Neutrino masses, mixings and vacuum 
oscillations have been studied for various
sizes of mass parameters.

In this paper we  continue to study the consequences of mixing of the 
usual neutrino with bulk fermions in the context  
elaborated in Ref. \cite{addm}. 
We consider both the neutrino oscillations (in vacuum and  medium) 
and the resonance conversion.  
We show that the resonance conversion 
of the electron neutrinos to the bulk states can solve 
 the solar neutrino problem.   This solution 
implies $R \sim 0.1$ mm, thus giving connection between neutrino
physics and sub-millimeter gravity measurements. 
We also discuss production of the KK- neutrino states in the Early
Universe and in supernovae.\\

\section{Neutrino mixing with the bulk modes. Universality}

Let us first assume that  extra dimensions  have the hierarchy of radii, 
so that only one extra dimension has radius $R$ in sub-millimeter
range and therefore only one tower of corresponding Kaluza-Klein modes is 
relevant for the low energy neutrino physics. 
The number and the size of other dimensions will be chosen to 
satisfy the constraint (\ref{radii}). (We will  comment on  effects 
of two sub-millimeter dimensions in sect. 4 and 5.)

The  right handed bulk states, $\nu_{i R}$, form with 
the left handed bulk components,   
$\nu_{i L}$, the Dirac mass terms  
which originate from the quantized internal momenta in extra dimension:    
\be
\sum_{n = -\infty}^{+\infty} m_n~\bar{\nu}_{n R} \nu_{n L} + h.c.,
~~~~~~~~~ 
m_n \equiv  \frac{n}{R}~. 
\label{mass}
\ee
The mass-split is determined  by $1/R$.  
According to (\ref{inter}, \ref{Dmass})
the bulk states  mix with usual  left handed neutrino
(for definiteness we will consider the electron neutrino
$\nu_{e L}$)  by the Dirac type mass terms   
with universal mass parameter: 
\be
m_D \sum_{n = -\infty}^{+\infty}  \bar{\nu}_{n R} \nu_{e L}, ~~~~~~~~~
m_D \equiv \frac{hv M_f}{M_P} \approx  
6 \cdot 10^{-5} {\rm eV} h  \frac{M_f}{1 {\rm TeV}}~,  
\label{mixing}
\ee
where $h$ is the renormalized Yukawa coupling.

The mass terms (\ref{mass},\ref{mixing}) can be rewritten as 
\be
m_D \bar{\nu}_{0 R}\nu_{e L} + 
m_D \sum_{n = 1}^{\infty} (\bar{\nu}_{n R} + \bar{\nu}_{- n R})\nu_{e L}
+ \sum_{n = 1}^{\infty} \frac{n}{R}~
(\bar{\nu}_{n R} \nu_{n L} - \bar{\nu}_{-n R} \nu_{-n L})  + h.c..
\label{mass1}
\ee
Notice that $\nu_{0L}$ decouples from the system. Introducing new states: 
\be 
\tilde \nu_{n L} = \frac{1}{\sqrt{2}}
(\nu_{n L} - \nu_{-n L})~, ~~~~~~~
\tilde \nu_{n R} = \frac{1}{\sqrt{2}}
({\nu}_{n R}  + {\nu}_{-n R})
\label{newstates} 
\ee
and denoting by $\nu'_{n L}$, $\nu'_{n R}$ the orthogonal 
combinations  we can write the mass terms in (\ref{mass}) as 
\be
m_D \bar{\nu}_{0 R}\nu_{e L} +  
\sqrt{2} m_D \sum_{n = 1}^{\infty} 
\bar{\tilde \nu}_{n R} \nu_{e L}
+ \sum_{n = 1}^{\infty} \frac{n}{R}~
\left(\bar{\tilde \nu}_{n R} \tilde \nu_{n L} + 
\bar{\nu}'_{n R} \nu'_{n L}\right)
+ h.c..
\label{mass2}
\ee
Notice that the zero mode has 
$\sqrt{2}$ smaller mixing mass  with  
$\nu_e$ than non-zero modes; the states  
$\nu'_{n L}$, $\nu'_{n R}$ decouple from the rest of 
the system.

Diagonalization of the mass matrix formed by the mass terms
(\ref{mass2}) in the limit of $ m_D R \ll 1$
gives (see Appendix)  the mixing of
neutrino with
$n^{th}$ - bulk mode, $\tilde \nu_{n L}$:
\be
\tan \theta_n \approx \frac{\sqrt{2} m_D}{m_n} =   \frac{\xi}{n}~,
\label{tan}
\ee
where
\be
\xi \equiv  \frac{\sqrt{2} h v M_f R}{ M_P}
\ee
determines  mixing with the first bulk mode.
The lightest state, $\nu_0$,  has the mass $m_0 \approx m_D $  and others,
$\tilde{\nu}_n$,: 
\be
m_n \approx \frac{n}{R}~.  
\ee

According to (\ref{tan})  the electron 
neutrino state can be written in terms of the mass eigenstates 
as
\be
\nu_e \approx \frac{1}{N} \left(\nu_0 + \xi \sum_{n = 1} \frac{1}{n}
\tilde{\nu}_{n}
\right)~,
\label{nue}
\ee
where the normalization factor $N$ equals
\be
N^2 = 1 +  \xi^2 \sum_{n = 1} \frac{1}{n^2} 
 = 1 + \frac{\pi^2}{6} \xi^2~.
\label{norm}
\ee

~From the phenomenological point of view the
bulk modes (being the singlets of the SM symmetry group)
can be considered as sterile neutrinos.
Thus, we deal with the coupled system of the  electron neutrino and
infinite number of sterile neutrinos mixed. According to (\ref{nue}), 
the electron neutrino
turns out to be  the coherent mixture of states with increasing mass and
decreasing mixing.\\

The following comment concerning normalization is in order.
The contribution 
to the normalization $N^2$ (\ref{norm}) from mass states with 
$n = k, k+1, ....$ (starting from number $k$)  can be estimated 
substituting the sum by the integral:
\be
\Delta_k \sim \xi^2 \int_k \frac{dn}{n^2} = \frac{\xi^2}{k}~
\label{norm}
\ee
and for $k \rightarrow \infty$ we get  $\Delta_k \rightarrow 0$.  
Thus, due to decrease of mixing the effect of heavy states is suppressed.

In real physical processes with energy release $Q$ only
low mass part of the state (\ref{nue}) can be produced.
The states with
$
n > Q R
$
do not appear. This  leads to breaking
of universality, that is,  to difference  of normalization of the 
neutrino states produced in  processes with different $Q$. 
As follows from (\ref{norm}) the contribution of states with $n > Q R$ to
the normalization equals
\be
\Delta (Q)  = \frac{\xi^2}{Q R} \approx  10^{-10}
\frac{h^2}{Q/1 {\rm MeV}}~.
\label{Qnorm}
\ee
(For $M_f = 10$ TeV and $R^{-1} = 3 \cdot 10^{-3}$ eV.) 
Since $Q \gg 1/R$ for sub-millimeter scale  the
deviation from
universality is negligible. This is not true for two 
extra dimensions of  common size  \cite{fp} (see later).
For the same reason a change of kinematics of
processes due to emission of
states with $m_n \sim Q$ is unobservable.
The probability of emission of the heavy states is
negligible. \\

\section{Oscillations and Resonance Conversion}

Let us consider the vacuum oscillations of neutrino state
produced as $\nu_e$ (\ref{nue}) to the bulk modes. 
The state will evolve with time $t$ as 
\be
\nu_e(t)  = \frac{1}{N} \left(\nu_0 + \xi \sum_{n = 1}
\frac{1}{n} e^{i\phi_n} \tilde{\nu}_{n}
\right)~,
\label{nuet}
\ee
where the phases $\phi_n$ equal
\be
\phi_n \approx \frac{(m_n^2  - m_D^2) t}{2 E}~, 
\ee
and  $E$ is the energy of neutrinos. The survival probability
of the  $\nu_e \leftrightarrow \nu_e$ oscillations then equals:
\be
P \equiv |\langle \nu_e |\nu_e(t) \rangle|^2
= \frac{1}{N^4}
\left| 1  + \xi^2 \sum_{n = 1} \frac{e^{i\phi_n}}{n^2} \right|^2.
\label{probab}
\ee
Since $\phi_n \propto n^2$, the oscillation picture  consists of 
interference of the infinite number of modes with increasing
frequencies $\propto n^2$ and decreasing amplitudes $\propto 1/n^2$.
For practical purpose all high frequency modes can be averaged, 
so that  only a few low frequency oscillations can be observed 
depending on the energy resolution of detector. Using
(\ref{probab}) we find the probability averaged over all the modes:
\be
\bar{P} = \frac{1}{[1 + (\pi^2/6) \xi^2]^2}
\left[ 1  + \frac{\pi^4}{90}\xi^4 \right].
\label{avera}
\ee
It is smaller than the  two neutrino probability   
with the same mixing parameter 
$\xi$ due to presence of infinite number of mixed
states.
In particular, for $\xi \ll 1$ we get $\bar{P} \approx 1 - (\pi^2/3)
\xi^2$,  
whereas in the $2\nu$ case: $\bar{P} \approx 1 - 2\xi^2$. 

In the case when only lowest frequency mode (associated to
$\nu_1$) is non-averaged, we get from  (\ref{probab})
the survival probability
\be
P = \frac{1}{(1 + (\pi^2/6) \xi^2)^2}
\left[ (1  + \xi^2)^2 +  \left(\frac{\pi^4}{90} - 1 \right)\xi^4
- 4 \xi^2 \sin^2 \frac{\phi_1}{2}
\right]~. 
\label{firstmode}
\ee
According to  this equation the  depth of oscillations
equals 
\be 
A_P =   \frac{4 \xi^2}{[1 + (\pi^2/6) \xi^2]^2}~. 
\label{depth}
\ee

Notice that due to presence of (practically) infinite number 
of the bulk modes which give just averaged oscillation result, 
the depth  of oscillations of the lowest mode can not be maximal.  
Moreover,  the relation between the depth and the average probability
differs from the standard 2$\nu$ - oscillation case (see also \cite{ddg}).

Similarly one can find the probability with two non-averaged modes,  
etc..\\

In medium with constant density the  mixing with
 different bulk states is  modified depending on 
$m_k^2$ and  the potential, $V$,   which describes  the interaction with
medium: 
\be
V  =  G_F \frac{\rho}{m_N} \left( Y_e - \frac{1}{2}Y_n \right)~.  
\label{potential}
\ee
Here $G_F$ is the Fermi coupling constant, $m_N$ is the nucleon mass,
$Y_e$ and  $Y_n$ are the numbers of electrons and neutrons per
nucleon correspondingly. 
The mode for which
the resonance condition \cite{MSW}, 
\be
\frac{m_k^2}{2E} \approx  V
\ee
is fulfilled (resonance modes)
will be enhanced: the effective mixing  will be enhanced  and 
the oscillations will proceed with large  depth. 
The modes with lower frequencies (masses) will be suppressed; 
the high frequency modes,  $m_k^2/2 E \gg  V$,  
will not be modified.\\

Let us consider propagation of neutrinos in medium with varying density
$\rho(r)$
keeping in mind  applications to solar and supernova  neutrinos. 
The energies of bulk states do not depend on density:
\be
H_i = \frac{m_i^2}{2E}~, 
\label{ilevel}
\ee
whereas for the electron neutrino we have
\be
H_e \approx   V(\rho) ~.  
\label{elevel}
\ee
Therefore the level crossing scheme (the ($H  - \rho$) -  plot which
shows the dependence of the
energies of levels $H_e, H_i$ on density)  consists of 
infinite number of horizontal parallel lines (\ref{ilevel})
crossed by  the electron neutrino line (\ref{elevel}). 
In what follows we will concentrate on the case 
$\xi \ll 1$, so that $m_D \ll m_n$ for all $n$,   
and  the crossings (resonances)  occur
in the neutrino channels.
The resonance density, $\rho_n$,  of  $H_e$ crossing with
energy of $n^{th}$ bulk state: $H_n = H_e(\rho_n)$
equals according to (\ref{ilevel}, \ref{elevel})
\be
\rho_n = \frac{m_n^2 m_N}{2 E G_F ( Y_e - \frac{1}{2}Y_n)} \propto n^2 ~.
\label{crossing}
\ee
For small mixing ($\xi \ll 1$ ) the resonance layers 
for different bulk states 
(where the transitions, mainly,  take place) 
are well separated:
\be
\rho_{n +1} - \rho_{n} \gg \Delta \rho_{n R} = \rho_n \frac{2
\xi}{n}~, 
\ee   
here $\Delta \rho_{n R}$ is the width of the $n^{th}$- resonance layer.  
Therefore  transformation in each resonance occurs independently
and the interference of effects from  different resonances can be
neglected.
In this case the survival probability $\nu_e \rightarrow \nu_e$
after crossing of $k$ resonances is just the product of the
survival probabilities in each resonance:
\be
P = P_1 \times P_2 \times ~ ....~\times  P_k~.
\ee
Moreover, for $P_i$ we can take the asymptotic formula   
which describes transition with 
initial density being  much larger than the resonance density
and the final density  --  much smaller than the resonance density.
As the first approximation we can use the Landau-Zenner formula
\cite{Parke}: 
\be
P_n \approx \left\{
\begin{array}{ll}
1 & E < E_{n  R} \\
{\displaystyle e^{- \frac{\pi}{2} \kappa_n}} & E > E_{n R}  
\end{array}
\right. ~, 
\label{LZ}
\ee
where  $E_{n R}$ is the resonance energy which corresponds
to maximal (initial) density $\rho_{max}$ in the region where neutrinos
are produced:
\be
E_{n R} \approx \frac{m_n^2 m_N}{2 E G_F \rho_{max} ( Y_e - \frac{1}{2}
Y_n)}~;
\label{renergy}   
\ee

\be
\kappa_n = \frac{m_n^2}{2E} \frac{\sin^2 2\theta_n}{\cos 2\theta_n}
\frac{\rho}{d\rho /dr}
\ee
is the adiabaticity parameter \cite{MSW} and 
$\sin^2 2\theta_n \approx 4\xi^2/n^2$. 
Since $m_n^2 \propto n^2$  whereas $\sin^2 2\theta_n \propto 1/n^2$, 
the adiabaticity parameter,  
$\kappa_n  = \kappa_1 \propto m_n^2 \sin^2 2\theta_n$,  
does not depend on $n$ for a given  energy.
Using this property we can write
final expression for the survival probability as
\be
P \approx
{\displaystyle e^{- \frac{\pi}{2} \kappa_1 f(E)}}~,
\label{surv}
\ee
where
\be
\kappa_1 \approx \frac{2 \xi^2}{E R^2} \frac{\rho}{d\rho /dr}~, 
\ee
and $f(E)$ is the step-like function
\be
f(E) =  \left\{
\begin{array}{ll}
0 , & E < E_{1 R} \\
n , &  E_{n R}  <  E < E_{n +1~R}
\end{array}
\right.~. 
\label{fff}  
\ee

Since high level resonances turn on at higher energies and
$\kappa \propto 1/E$, the effect of conversion decreases with
increase of the order of the resonance, $n$. Moreover, 
since in real situation the density is restricted from above 
and the energies of neutrinos are in certain range,
only finite number of levels is relevant 
and  the largest effect is due to the lowest mass resonance.\\
   
\section{Solution of the Solar Neutrino Problem}

Let us apply the results of previous section to
solution of the solar neutrino problem.

We choose the lowest (non-zero) bulk mass, $m_1 = 1/R$, in such a way
that $\Delta m^2 = 1/R^2$ is in the range of small mixing
MSW solution  due to conversion to sterile neutrino
$\nu_e \rightarrow \nu_s$ \cite{BKS}:
\be
\frac{1}{R^2} = (4 - 10) \cdot 10^{-6}~ {\rm eV}^2~.
\label{masssq}
\ee
This  corresponds to $1/R = (2 - 3)~10^{-3}$ eV or
\be
R = 0.06 - 0.10~ {\rm mm}~.
\ee
Using mass squared difference (\ref{masssq}) as well as  
maximal density and chemical
composition of the sun we 
find from (\ref{renergy}) 
\be
E_{1R} \approx 0.4 \div 0.8~ {\rm MeV}~. 
\ee
Thus  the $pp$-neutrinos ($E_{pp} < 0.42$  MeV) 
do not undergo resonance conversion: $E_{pp} < E_{1 R}$ 
(see (\ref{surv}, \ref{fff})), 
whereas the beryllium neutrinos ($E_{Be} = 0.86$ MeV) 
cross the first resonance. 
(For smaller $1/R^2$ the 
$pp$-neutrinos from the high energy part of their spectrum 
can cross the resonance and  the flux can be partly 
suppressed.)

The energies of the next resonances equal:
$
E_{n R} = n^2 E_{1 R}, 
$
or explicitly:
$E_{2 R} = 1.6 - 3.2$ MeV,
$E_{3 R} = 3.6 - 7.2$ MeV,
$E_{4 R} = 6.4 - 12.8$ MeV,
$E_{5 R} =  10 - 20 $ MeV,
 $E_{6 R} =  14.4  - 28.8 $ MeV. Higher resonances ($n > 6$) turn on at
energies higher than maximal energy of the solar neutrino spectrum and
therefore are irrelevant.  The dependence of the
survival probability on energy is shown in the fig. 1.
The dips  of the survival probability 
at the energies $\sim E_{i R}$   
are due   to turning the  corresponding resonances.

The effects of higher resonances lead to additional 
suppression of the survival probability   in comparison with 
the two neutrino case. Therefore the
parameter $4\xi^2$ which is equivalent to $\sin^2 2\theta$ should be
smaller. We find that
\be
4\xi^2 = (0.7 - 1.5)\cdot 10^{-3}
\label{xival}
\ee
gives average suppression of the boron neutrino flux required by the
SuperKamiokande results. 
According to (\ref{Dmass}), the value of $\xi$ (\ref{xival})
determines the fundamental scale:
\be
M_f = \frac{\xi M_P}{\sqrt{2} hv R} = \frac{1}{h} (0.35 - 0.7)~ {\rm
TeV}~. 
\label{scale}
\ee
For small $h$ the scale $M_f$ can be large enough to satisfy 
various phenomenological bounds. 

Let us consider features of the suggested solution of the solar neutrino
problem.    
The solution gives the fit of total rates in all experiments  as good as
usual  2$\nu$ flavor conversion does: the  $pp$-neutrino flux is
unchanged  
or weakly suppressed, the beryllium neutrino flux can be strongly
suppressed,
whereas the boron neutrino flux is moderately suppressed and
this latter suppression  can be tuned by small variations
of $\xi$.

\begin{figure}[htb]
\hbox to \hsize{\hfil\epsfxsize=11cm\epsfbox{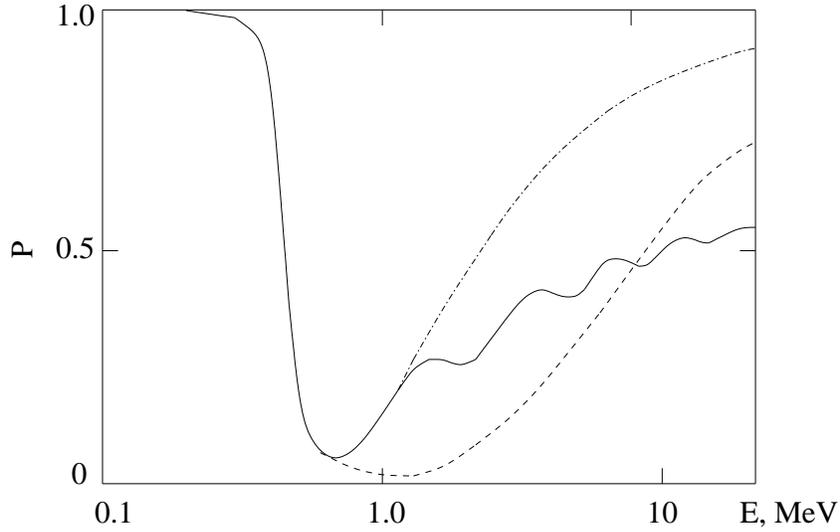}\hfil}
\caption{~~The survival probability as the function of neutrino 
energy for the electron neutrino conversion to the bulk 
states in the Sun (solid line), $4\xi^2 = 10^{-3}$. 
Dot-dashed line shows the survival probability 
of the  two neutrino conversion for the equivalent mixing  
$\sin^2 2\theta  = 10^{-3}$. Dashed line 
corresponds to the survival probability of the 
 two neutrino conversion for $\sin^2 2\theta  = 4 \cdot 10^{-3}$ 
which gives good fit of the total rates in all experiments. 
}
\label{fbimaxinv}
\end{figure}

Novel feature appears in distortion of the boron neutrino spectrum.
As follows from fig. 1 three resonances turn on in the
energy interval accessible by SuperKamiokande ($E > 5$  MeV).
The resonances lead to the wave-like modulation of the
neutrino spectrum. (Sharp form (\ref{surv}, \ref{fff}) 
is smeared due to  integration over 
the production region.)  The observation of such a regular wave structure   
with  $E \propto n^2$ would be an  evidence  of the
extra dimensions. However, in practice this will be very difficult to 
realize.  

The SuperKamiokande  experiment measures the energy spectrum 
of the recoil electrons from the reaction $\nu~ e - \nu~ e$ \cite{SK}. 
Integrations over the  neutrino energy as well as  the 
electron energy of the survival probability folded with the 
neutrino cross-section and the 
electron energy resolution function lead to strong  
smearing of the distortion 
in  the recoil spectrum. As the result,  the electron energy 
spectrum will have just small positive slope (the larger the energy 
the weaker suppression) 
with very weak ( $ < 2 - 3$ \%)  modulations.   It is impossible to
observe such
a modulations with present statistics. 

Notice that relative modulations become stronger if mixing,  
$\xi$, is larger than $10^{-3}$ and 
therefore  suppression is stronger. This, however,  requires larger 
original boron neutrino flux.
The SNO experiment \cite{SNO}  will have better sensitivity to
distortion of the spectrum. 

The slope of distortion of the neutrino spectrum is substantially smaller
than in the 
case of conversion to one sterile neutrino (see fig. 1). 
In view of smearing effects due to integration over neutrino and electron
energies (due to finite energy resolution)  
we can approximate the step-like function 
$f(E)$ in (\ref{fff}) by smooth function $f_{app}(E) \approx 
\sqrt{E/E_{1R}}$.  
Then  the smeared survival probability in the high energy range 
can be written as  
\be 
P \approx {\displaystyle e^{-\sqrt{\frac{E_0}{E}}}}~, 
\label{pro}
\ee 
where $\sqrt{E_0} = \pi \xi^2 \rho /(R^2 d\rho/dr \sqrt{E_{1R}})$. 

In the case of two large dimensions with 
$R_1 \sim R_2 \sim 0.02 - 0.03~ $ mm (see sect. 5) the number of 
bulk states, and therefore the number of relevant resonances
increases quadratically: $n^2$. (Here $n \sim 5 - 6$ 
is the number of resonances in the energy range of solar neutrinos 
in the one dimension case.) Correspondingly, the approximating function 
$f_{app}(E)$ will be proportional  to $E$. 
As the result, $\kappa_1 \cdot f(E) = const$  
and the smeared survival probability 
will not depend on energy. In this case $P(E)  \approx const$ for 
$E > E_{1R}$ and there is no distortion of the
recoil electron spectrum.  For two different radii: $R_2 < R_1$
one can get any intermediate behaviour of the probability between 
that in (\ref{pro}) and $P = const$.

Common signature of both standard $\nu_e - \nu_s$ conversion and
conversion to the bulk modes is the suppression of the neutral current
(NC) interactions. The two can be, however,  distinguished using the
following fact. 
In the case of the $\nu_e - \nu_s$ conversion there is certain correlation
between suppression of the NC interactions and distortion of the spectrum. 
The weaker distortion the  weaker suppression of the 
NC interactions and vice versa. In the case of $\nu_e - \nu_{bulk}$ 
conversion a weak distortion can be accompanied by significant
suppression of the NC events.  
This can be tested in the  SNO experiment.

No significant Day-Night asymmetry is expected due to
smallness of mixing angle. 

Thus, the smeared distortion of the energy spectrum 
(for $E > 5$ MeV ) is weak or absent in the case of 
$\nu_e - \nu_{bulk}$ conversion. However, in contrast to other energy
independent solutions here $pp-$neutrino flux may not be 
suppressed, or the energy spectrum of $pp$- neutrinos can be 
significantly distorted.\\

Notice that it is impossible to reproduce the large mixing 
angle MSW solution of the solar neutrino problem \cite{BKS} in 
this context. Indeed, for $\xi \sim O(1)$ and 
$1/R^2$ as in (\ref{masssq})  transitions in all low mass resonances 
are  adiabatic, and therefore the survival probability 
has the form: 
\be
P(E) \sim P_n(E) = \sin^2 \theta_n \approx \frac{\xi^2}{n^2}, ~~~~
E_{n R} < E < E_{n +1~R}.  
\label{lma}
\ee
(There are smooth transitions in the regions $E \sim E_{n R}$.) 
For a given energy $E$ the probability  is determined by 
conversion in the nearest $n^{th}$ resonance with  
$E_{n R} < E$.  According to (\ref{lma}) 
the probability   decreases 
monotonously with energy, in
contrast with observations. For instance, if $P \sim 0.5$ 
in the interval 
$E = 0.5 \div 2$ MeV, then it will be 
$P \sim 0.2$  for $E = 2 \div 4.5 $ MeV, 
 $P \sim 0.06$ for $E = 4.5 \div 8$ MeV etc.. 

If $\xi > 1$,  for all the modes $k$  with $k^2/(k +1) < \xi^2$ 
the resonances will be in the antineutrino channels and 
for $k^2/(k +1) > \xi^2$ in the neutrino channels \cite{DS}. \\

Note that solution of  the solar neutrino problem
due to the long length vacuum oscillations \cite{VO} into 
bulk modes  implies  
$\Delta m^2 \approx (0.5 - 5) \cdot 10^{-10}$ eV$^2$ 
and large (maximal) mixing. This leads to the following estimations
\be 
\frac{1}{R} \sim  \sqrt{\Delta m^2} \sim (0.7 - 2)~ 10^{-5} {\rm eV},~~ 
m_D = \frac{\xi}{R} = (0.5 - 1.5) \cdot 10^{-5} {\rm eV} 
\ee
for  values $\xi =  0.5 - 0.7 $. 
Now the size of the extra dimension equals $L = 6 - 20$ cm   
which is  excluded by existing tests of the Newton law.\\ 

The approach opens however another possibility. Suppose that 
the radius $R$ is small enough so that KK-excitations 
have negligible mixing with usual neutrinos. In this case 
the effect of extra dimensions is reduced to 
interaction with zero mode,  
$\nu_{0R}$, only. Suppose  that the
same bulk field 
couples with two usual  neutrinos: 
$\nu_e$ and $\nu_{\mu}$ (or $\nu_{\tau}$) generating the 
Dirac mass terms 
\be
m_{eD} \bar{\nu}_{0R} \nu_e + m_{\mu D} \bar{\nu}_{0R} \nu_{\mu}. 
\ee
These  terms lead to Dirac neutrino with mass 
$m_D = \sqrt{m_e^2 + m_{\mu}^2}$ formed by ${\nu}_{0R}$ 
and the combination 
$(m_{eD} \nu_e + m_{\mu D} \nu_{\mu})/m_D$. The orthogonal 
component is massless. In this way the $\nu_e$ and $\nu_{\mu}$ 
turn out to be mixed with the  angle determined 
by $\tan \theta = m_{eD}/m_{\mu D}$. 
(Similar mechanism of mixing has been considered in Ref. \cite{ddg}.) 
For $M_f \sim 1$ TeV and 
the original Yukawa couplings with bulk field 
$h_e \sim h_{\mu} \sim1 $ we get  $m_{eD} \sim m_{\mu D} \sim 
10^{-5}$ eV which leads to 
$\Delta m^2 \sim 10^{-10}$ eV$^2$ and maximal (large) 
$\nu_e - \nu_{\mu}$ mixing. This reproduces values of parameters 
required for the  
$\nu_e \leftrightarrow \nu_{\mu}$ Just-so oscillation solution of the 
solar neutrino problem. The solution has however no 
generic signatures of the high dimensional theory 
and employs the latter as the source of neutrino mass only.

\section{Solar Neutrinos and Parameters of Extra Dimensions}

As follows from previous section,  a solution of the solar neutrino
problem via resonance conversion to the bulk modes implies 
the radius of  extra dimension   $R = 0.06 - 0.1$ mm and  
the fundamental scale $M_f > 0.5 - 1$ TeV. To satisfy the relation
(\ref{radii}) we need to introduce  more extra dimensions.
Let us assume that second  large  
dimension exists with radius $R'$. From (\ref{radii}) we get 
\be 
\frac{1}{R'} =  \frac{(2\pi)^2 M_f^4 R }{M_P^2} \approx 
1.3 \cdot 10^{-3} \left(\frac{M_f}{1 {\rm TeV}}\right)~{\rm eV}.  
\ee
For $M_f = 1 $ TeV both  extra dimensions will have 
radii of the same size. In this case more  bulk states are involved in 
conversion of solar neutrinos which will lead to absence of the 
distortion of the spectrum, as we have discussed in sect. 4.

For $M_f \geq 10 $ TeV the bulk states associated  to $1/R'$ 
dimension do  not participate in solar neutrino conversion, 
however they are relevant for other 
neutrino processes. Let us consider this in more details. 
Now the electron neutrino state can be written as 
\be
\nu_e = \frac{1}{N} \left(\nu_0 + 
\xi \sum_{n} \frac{1}{n} \nu_{n,0} +
\xi \sum_{n, k \geq 1}
\frac{1}{\sqrt{n^2 + (R/R')^2 k^2}}
\nu_{n, k}
\right)~,
\label{nue21}
\ee
where index $n$ refers to dimension of larger radius  $R$, 
while  $k$   enumerate the bulk  states from  
dimension  $R'$. The sum over the states 
is  divided into two parts: the first sum contains the states 
with $k = 0$, that is, with small mass split only. This part corresponds
to  the one dimensional case discussed in sect. 2 - 4 and 
as we have shown in sect. 3,  
it does not lead to observable violation of universality.  
The second sum contains the towers of states with  
both large  and small  mass splits. 
Its contribution to the normalization of the state equals 
\be
\Delta  \equiv
\xi^2 \sum_{n, k} \frac{1}{n^2 + (\frac{R}{R'})^2 k^2}~.
\label{univ}
\ee
We can estimate the sum substituting it by the integral 
over $n$ and $k$. Performing first integration over 
$n$ from 0 to $\infty$ and then over $k$ from 1 to $(Q R')$ -- 
the number of states which can be produced in the process with energy
release $Q$,   we find: 
\be
\Delta (Q)  = 
\frac{\pi}{2} \frac{R'}{R} \xi^2 \ln (Q R') = 
\pi 
\left(\frac{h v}{M_P}\right)^2 \frac{V_2 M_f^2}{(2\pi)^2} \ln (Q R')~. 
\label{univ1}
\ee
The difference of normalizations of the two states produced in 
processes with energy releases 
$Q_1$ and $Q_2$ gives  measure of the universality breaking: 
\be   \Delta_{21} \equiv  \Delta (Q_2) - 
\Delta (Q_1)  =  
\frac{\pi}{2} \frac{R'}{R} \xi^2 \ln (Q_2/Q_1). 
\ee
Taking $\xi^2 = (2 - 4) \cdot 10^{-4}$ as is implied by the solar neutrino
data, $Q_1\sim 1$ MeV and $Q_2 \sim 100$ GeV as well as $R'/R < 0.1$ 
we find $\Delta_{12} \sim (2 - 4) \cdot 10^{-4}$ which is below 
present sensitivity. Notice, however,  that for $R' \sim R$  
the violation can be at the level of existing bounds \cite{fp}. \\

Let us consider the case of three extra dimensions with 
radii $R \gg R_2,  R_3$.  
Assuming for simplicity the equality  $R_2 = R_3  \equiv  R'$,  
we find from (\ref{radii}):  
\be
\frac{1}{R'} =  \frac{(2\pi)^{3/2}M_f^2 \sqrt{M_f R }}{M_p}~.
\ee
For $1/R \sim 2.5 \cdot 10^{-3}$ eV and $M_f = 10~$ TeV this equation
gives  
$1/R' \sim 3 \cdot 10^{-2}$ GeV.  
Performing calculations similar to those for one additional dimension we
find parameter of universality violation 
\be 
\Delta_{21} \approx 2 \left(\frac{h v}{M_P}\right)^2 
\frac{V_3 M_f^3}{(2\pi)^3} \frac{(Q_2 - Q_1)}{M_f} = 
\frac{(h v)^2 (Q_2 - Q_1)}{M_f^3} 
\ee
($V_3 = (2\pi)^3R R'^2$) 
which can be as low as  $10^{-6}$ 
even for $Q_2 - Q_1 \sim 100$ GeV 
(in this estimation we used  $M_f = 10$ TeV and $h = 1$).\\

\section{Astrophysical Bounds. Atmospheric neutrinos}

The important bound on mixing of usual neutrinos with 
sterile neutrinos as well 
as with bulk states follows  from primordial nucleosynthesis: 
production of new relativistic degrees of freedom 
leads to faster expansion of the Universe and to larger 
abundance of $^4 He$. There are two ways of production of the bulk states: 
(i) via oscillations  and (ii) incoherently via chirality flip. 
 
In the first case  the electron neutrino produced as the coherent 
combination of mass eigenstates  oscillates in medium 
to bulk states. Inelastic collision splits the state into 
active ($\nu_e$) and to sterile ($\nu_{bulk}$) parts 
and after each collision 
two parts will oscillate independently. 
Oscillations  between two collisions  average. Production 
rate is then 
the  sum of the averaged oscillation effects over collisions. 
The condition that sterile states do not reach equilibrium 
puts stringent bound on oscillation parameters  \cite{NS}.

The masses squared and mixing angles 
of bulk states (\ref{masssq}, \ref{xival}) 
implied by  solution of the solar neutrino problem   
satisfy the following relation: 
\be
\Delta m^2_n \cdot \sin^2 2\theta_n \approx \frac{4\xi^4}{R^2}~= 
(4 - 8)\cdot 10^{-9} {\rm eV}^2,  
\ee 
where $\sin^2 2\theta_n \equiv 4\xi^4/n^2$. 
This ``trajectory" in the 
$\Delta m^2  - \sin^2 2\theta$ - plot 
lies far outside  the region  excluded by primordial 
nucleosynthesis \cite{NS}. 
That is, production of the bulk neutrinos 
via oscillations is strongly suppressed. 

Let us consider  incoherent production of the bulk states due to  
chirality flip. The  production rate of
an individual bulk neutrino ($\Gamma_{1}$) is suppressed   
relatively to the production rate 
of the left-handed neutrino ($\Gamma_{\nu_e}$) by the chirality - flip
factor  
\be
{\Gamma_{1}\over \Gamma_{\nu_e}} \sim \left ({m_D \over T}\right)^2~, 
\ee
where the temperature $T$  in the denominator comes from the
propagator of a primarily-produced left handed  neutrino. The 
multiplicity of the final bulk states is 
$T R$,  so that the total bulk neutrino  emission rate is  
suppressed as 
\be
{\Gamma_{bulk}\over \Gamma_{\nu_e}} \sim 
\left ({m_D \over T}\right)^2~ T R = \xi \left ({m_D \over T}\right)~. 
\label{ratio}
\ee
For the parameters implied by the solar neutrinos and $T \sim 1$ MeV 
this ratio is about $3 \cdot 10^{-11}$. 
Using (\ref{ratio}) we find the temperature, $T_{bulk}$,  
at which 
production rate of the bulk states is comparable with expansion rate 
of the Universe: $\Gamma_{bulk} = \Gamma_{exp} \sim T^2/M_P$: 
\be
T_{bulk} = T_{\nu} \sqrt{\frac{T_{\nu}}{\xi m_D}}~, 
\label{tempbulk}
\ee
where $T_{\nu} \sim 1$ MeV is the temperature 
of the neutrino decoupling. From (\ref{tempbulk}) 
we find $T_{bulk} \sim  200$
GeV which is much above the ``normalcy" temperature \cite{add1}.    
The modes from additional dimensions having smaller radii 
give even smaller contribution. \\

The KK-neutrinos as well as  the   KK-gravitons 
produced in  stars, in particular, in  supernovae, increase  the rate of
star cooling \cite{add1}. 
No additional sources of cooling have been found from observations of 
the SN1987A which put stringent bound on production of the bulk states. 
The rate of the incoherent production of the bulk neutrinos in supernovae 
is suppressed by the same factor (\ref{ratio}). 
For  temperature of  the core of supernova $T \sim 30 $ MeV 
the eq. (\ref{ratio}) gives 
$\Gamma_{bulk}/ \Gamma_{\nu_e} \sim 3 \cdot 10^{-12}$. 
Then taking into account that bulk neutrinos are emitted from the whole 
volume of the core, whereas  usual neutrinos are emitted from the 
surface (neutrinosphere) we find  
that the luminosity in the bulk states is 5 - 6 orders of magnitude smaller
than luminosity in neutrinos. 
Production of the bulk states via oscillations is strongly suppressed 
by matter effect. Matter suppression is weak or absent for 
bulk states with high mass: $m \sim 10^3$ eV. However, their production 
is suppressed  by very small vacuum mixing.

So, we conclude that  parameters required by 
solution of the solar neutrino problem satisfy 
astrophysical constraints.\\

Let us comment in this connection on solution of the atmospheric neutrino
problem via oscillations of muon neutrinos to the bulk states 
$\nu_{\mu} \leftrightarrow \nu_{bulk}$. This solution 
requires smaller radius of the extra dimension: 
$1/R \approx \sqrt{\Delta m^2_{atm}} \sim (5 - 9)\cdot 10^{-2}$ eV 
or  $R \sim (2 - 4)\cdot 10^{-3}$ mm and near to maximal mixing: 
$\xi \sim 1$. The latter means that the fundamental scale should be 
about $M_f \approx (10^3~ {\rm TeV})/ h$. 
that is,  about 3 orders of magnitude larger than that
for solar neutrinos. 
(Notice that  approximation $\xi \ll 1$ can not be used 
now to diagonalize the mass matrix and results 
of sect. 3 and Appendix should be corrected \cite{DS}.   
Still mixing of large mass bulk states is suppressed by factor 
$1/n$ and these states lead to finite averaged oscillation result. 
Only a few low mass states are relevant for non-averaged oscillation
picture.) 

The oscillation parameters required by the  solution of the 
atmospheric neutrino problem violate nucleosynthesis bound. 
Indeed,  masses and mixing angles  of the bulk modes satisfy now the
relation: 
\be
m_n^2 \cdot \sin^2 2\theta_n \sim \frac{2}{R^2} \sim 6 \cdot 10^{-3} 
{\rm eV}^2. 
\ee
Nucleosynthesis bound reads \cite{NS}  
\be
\Delta m^2 < \frac{3 \cdot 10^{-5} {\rm eV}^2 } {\sin^4 2\theta}~.  
\ee
{}From these two equation we find that about 20 - 25 lightest 
bulk states are in the forbidden region. They turn out to  
be in equilibrium, whereas 1 or at most  2 are allowed. 
Production of the bulk states via oscillations can be suppressed if there
is substantial ($\sim 10^{-5}$) leptonic asymmetry in the Universe 
\cite{FV}.\\

Let us finally consider production of gravitons in the stars. 
The generic reason that 
saves the bulk gravitons from being ruled out by star cooling 
is the infrared-softness of the high dimensional gravity. 
On the language of four-dimensional KK modes this  can be
visualized as 
follows. The rate of each individual KK graviton emission in the star
is suppressed by the  universal volume factor
\be
\left ({T \over M_P} \right)^2 \sim \left ({T \over M_{f}}
\right)^2 
{1  \over M_f^N V_N}~. 
\ee
The number of available final states is $ \sim T^NV_N$,
so that the over-all rate is suppressed as 
\be
\Gamma_{grav} \sim  \left({T \over M_P} \right)^2 T^N V_N 
 \sim \left ({T \over M_f}
\right)^{2 + N}~. 
\label{supern}
\ee
According to this expression, the analysis of supernova core 
cooling 
gives for $N = 2$ the lower bound on the fundamental gravitational
scale $M_f$
about  $30$ TeV \cite{add1} - $50$ TeV \cite{supernova22} .
As it is clear   from (\ref{supern}), the rate is determined
by the value of $M_{f}$ and  it  is insensitive to 
sizes of individual  radii $R_i$ as far as all $1/R_i < T$.
On the other hand,  if some radii, $R_k$, do not satisfy this bound, then
the
corresponding KK modes can not be produced in the star and
the corresponding factor $T R_k$ in $T^NV_N$ of Eq. (\ref{supern}) has 
to be replaced by 1.
The bottom-line of this discussion is that standard constraint can be
avoided if the extra dimensions have different radii. 
For instance, with one 
radius $R \sim 0.03$ mm   and two others $R_k > T$ 
(which can be realized even for 
$M_f$ as low as several TeV's)  the rate becomes
\be
\Gamma_{grav} \sim \left ({T \over M_P} \right)^2 (T R) 
\approx 10^{-31}~.  
\ee
Therefore, some of the radii can be of sub-millimeter size and,
thus, can be a subject of  direct experimental search in
future gravitational measurements \cite{measurements}.\\

\section{Conclusions}

We have studied consequences of the neutrino mixing with  fermions
propagating in the bulk  in the context of theories with 
large extra dimensions. The bulk fermions could be 
components of bulk gravitino or other singlets of the SM gauge
group. 

Phenomenology of this mixing is determined by the following  features: 
(i) The bulk fermions 
can be considered as sterile neutrinos.   
(ii) Large  number of these sterile neutrinos is 
involved in physical processes.  
(iii) For $m_D < 1/R$  usual neutrinos are  combinations of 
mass eigenstates with increasing masses and decreasing admixtures.

The effect  of  bulk states with large masses is reduced to 
averaged oscillations. Low modes can lead to 
non-averaged oscillations in vacuum (uniform medium)  
or to multi-resonance conversion 
in medium with varying density.

The resonance conversion of the electron neutrino 
to the bulk states can solve  
the solar neutrino problem. Properties of this solution 
are similar  to those due to  conversion to sterile neutrino. 
The  important difference is that significant suppression of the 
boron neutrino flux can be accompanied by weak distortion of the 
energy spectrum. Moreover, weak  modulation of the boron neutrino 
spectrum is
expected due to conversion to several KK-states. 

Simultaneous explanation of the atmospheric, solar
and  LSND results in terms of neutrino oscillation/conversion 
implies existence of sterile neutrino. Moreover, the data 
favour  $\nu_{\mu} \leftrightarrow \nu_{\tau}$ 
oscillations as the  solution of  the atmospheric neutrino problem,  
so that  the solar neutrinos should be converted 
to sterile states. In this connection 
one can consider the following possibility. 
There is some  usual (4 dimensional) mechanism of generation 
of the  active neutrino masses. 
This mechanism  produces  neutrino 
mass pattern  with heavy and strongly mixed $\nu_{\mu}$  and 
$\nu_{\tau}$ and very light $\nu_e$. Such a pattern 
can explain the atmospheric neutrino  and  LSND results. 
Neutrinos 
(in general  of all flavors) couple also with the bulk fermion.    
These couplings (being of the same order for all neutrino species)    
generate  negligible mixing of the KK-states with heavy 
$\nu_{\mu}$ and $\nu_{\tau}$ and large enough mixing 
with light $\nu_e$,  so that the solar neutrino problem can be solved as 
is described in sect. 4.\\

The suggested solution of the solar neutrino problem implies that 
the radius  of at least one extra dimension is in the range 
0.06 - 0.10 mm, that is, in the range of sensitivity of 
proposed gravitational measurement.   
The fundamental scale should be about  
$h M_f \sim 1$ TeV. This mass satisfies the fixed overall
volume condition provided  additional large extra dimensions 
exist. In the case of one additional extra dimension its radius should be 
$1/R' = 1.3 \cdot 10^{-3} (M_f / 1 {\rm TeV})^4$ eV. 
For $h \sim 1$  one has  $ M_f \sim 1$ TeV,  
and $1/R' \sim 2 \cdot 10^{-3}$ eV,  so that the  
second extra dimension will  influence the solar neutrino
data too. If $h \ll 1$, the fundamental scale  can be much larger 
than 1 TeV, and 
$R'$ can be  much smaller than 1 mm. For $h = 0.1$ and $M_f = 10$ TeV we
get $1/R' \sim 10$ eV. For two additional dimensions  
the common radius equals $1/R' \sim 3 \cdot 10^{-2}$ GeV, 
if $h = 1$ and $M_f = 10$ TeV. 

For large fundamental scales ($M_{f}   > 10  - 20$ TeV),  
direct laboratory searches at high energies will be 
practically impossible and  
neutrinos can give unique (complementary to gravitational measurements)
opportunity  to probe the  effects of large extra dimensions.\\

\noindent
{\Large \bf Appendix}\\

The mass terms (\ref{mass2}) can be written as 
$$
\bar{\nu}_L M \nu_R~,   
$$ 
where ${\nu}_L^T \equiv (\nu_L, \tilde{\nu}_{1 L}, 
\tilde{\nu}_{2 L} ... )$ and 
$\nu_R^T =  (\nu_{0 R}, \tilde{\nu}_{1 R}, \tilde{\nu}_{2 R} ... )$; 
the  modes 
$\nu_{0 L}$, $\nu'_{n L}$ $\nu'_{n R}$
 decouple  from the system, and the mass matrix 
$M$ for  $k+1$ states  equals 
$$
M = \left(
\begin{array}{ccccc}
m_D & \sqrt{2}m_D & \sqrt{2}m_D          & ...   &\sqrt{2}m_D \\
0   & \frac{1}{R}  &  0  &  ...  & 0   \\
0   & 0   &  \frac{2}{R} & ...   & 0   \\ 
... & ...  & ...         & ...   & ...  \\
0   & 0    & 0           & ...   & \frac{k}{R} 
\end{array}
\right)~.  
$$
Let us consider the matrix $M M^{\dagger}$ which determines 
mixing of the left handed neutrinos: 
$$
~~~~~~~~~~~~~~~~~~~M M^{\dagger} = \frac{1}{R^2} \left(
\begin{array}{ccccc}
(k + 1/2) \xi^2 & \xi  & 2 \xi  & ...   &  k \xi \\
\xi    & {1}  &  0  &  ...  & 0   \\
2 \xi   & 0   &  4 & ...   & 0   \\
... & ...  & ...         & ...   & ...  \\
k \xi   & 0    & 0           & ...   & k^2
\end{array}
\right)~, ~~~~~~~~~~~~~~~~~~~~~~~~~~ (A1)
$$
where $\xi \equiv \sqrt{2} m_D R$.  
Notice that the mass matrix (A1)  
corresponds to compactification on the circle; 
it can be shown that the same matrix follows from 
the $Z_2$ - orbifold compactification \cite{ddg}. 
The diagonalization of matrix can be performed starting 
from the heaviest state $k \gg 1$. Then one can check that  
the limit $k \rightarrow \infty$ does not change results. 
The rotation by the angle $\theta_k$ in the plane $\nu_{0 L} - 
\tilde{\nu}_{k L}$
$$
~~~~~~~~~~~~~~~~~~~~~~~~~~~~~~~~\tan 2 \theta_k = 2 \frac{\xi}{k}\cdot 
\frac{1}{1 - \frac{\xi^2}{k}}
~~~~~~~~~~~~~~~~~~~~~~~~~~~~~~~~~~~~~~(A2)
$$
diagonalizes corresponding submatrix. 
It is easy to perform the diagonalization using $\xi$ as an expansion 
parameter:  $\xi \ll 1$ as is implied by the 
solution of the solar neutrino problem.  
The rotation (A1) leads to modification of the 
first diagonal element 
$
(k + 1/2) \xi^2 \rightarrow  (k -1/2) \xi^2 + O (\xi^4) 
$ 
and modification of the  mixing terms as 
$n \xi \rightarrow \cos \theta_k n \xi$. 
For small $\xi$: $\tan \theta_k \approx \xi/k$ and 
the  eigenvalues equal $ m_k^2 \approx k^2/R^2$.   
After $k-1$ subsequent rotations  we get for the first 
diagonal element: $3/2 \xi^2 + O (\xi^4)$ 
and for off-diagonal term: 
$$ 
\xi \cos \theta_2\cos \theta_3 ... \cos \theta_k \approx 
\xi \left(1 - \frac{1}{2}\xi^2 \sum_{n = 2 ... k} 1/n^2 +...
\right) \approx \xi. 
$$
\\
These results can be obtained from the exact characteristic equation 
for the mass eigenstates: ${\rm Det}[ M M^{\dagger} - m^2]$ which 
can be written explicitly as 
$$
\pi \xi^2 \cot(\pi m R) = 2 mR.  
$$
which coincides with the characteristic equation in \cite{ddg} for the 
same neutrino system.\\

\noindent
{\Large \bf  Acknowledgments}\\

One of us (A.S.) is grateful  to E. Dudas  for useful discussions.


\end{document}